\documentclass[reprint,amsmath,amssymb,aip,jap,floatfix]{revtex4-1}
\usepackage{graphicx}
\usepackage{dcolumn}
\usepackage{bm}
\usepackage{multirow}
\usepackage{color}
\usepackage[english]{babel}
\usepackage[T1]{fontenc}
\usepackage[caption=false]{subfig}
\usepackage{amsmath}
\usepackage{amsfonts}
\usepackage{amssymb}
\usepackage{color}
\usepackage{tabularx,booktabs}
\newcolumntype{Y}{>{\centering\arraybackslash}X}
\usepackage{dcolumn}
\usepackage{natmove}
\usepackage[unicode=true, bookmarks=true, bookmarksnumbered=false, bookmarksopen=false,
  breaklinks=false,pdfborder={0 0 0},backref=false,colorlinks=true]{hyperref}
 \hypersetup{pdftitle={Topologically nontrivial electronic states in superconducting CaSn$_3$},
  pdfauthor={Rinkle Juneja, Sunny Gupta, Ravindra Shinde, Abhishek K. Singh},
  pdfsubject={Computational Material Science},
  unicode=false,pdftoolbar=true,pdfmenubar=true,pdffitwindow=true,pdfstartview={FitH},pdfnewwindow=true,pdfcreator={RevTeX},linkcolor=red,citecolor=blue,urlcolor=blue}
  
\begin{document}

%%%%%%%%%%%%%%%%%%%%%%%%%%%%%%%%%%%%%%%%%%%%%%%%%%%%%%%%%%%%%%%%%%%%%%%%%%
%% Title, author, affiliation and date
%%%%%%%%%%%%%%%%%%%%%%%%%%%%%%%%%%%%%%%%%%%%%%%%%%%%%%%%%%%%%%%%%%%%%%%%%%
\title{Topologically nontrivial electronic states in CaSn$_3$}% Force line breaks with \\
\author{Sunny Gupta}
\author{Rinkle Juneja}
\author{Ravindra Shinde}
\author{Abhishek K. Singh}
\email{abhishek@mrc.iisc.ernet.in}
\affiliation{Materials Research Center, Indian Institute of Science, Bangalore - 560012, India}
\date{\today}

%%%%%%%%%%%%%%%%%%%%%%%%%%%%%%%%%%%%%%%%%%%%%%%%%%%%%%%%%%%%%%%%%%%%%%%%%%
%% ABSTRACT
%%%%%%%%%%%%%%%%%%%%%%%%%%%%%%%%%%%%%%%%%%%%%%%%%%%%%%%%%%%%%%%%%%%%%%%%%%
\begin{abstract}
Based on the first-principles calculations, we theoretically propose topologically non-trivial states in a recently experimentally discovered superconducting material CaSn$_3$. When the spin-orbit coupling (SOC) is ignored, the material is a host to three-dimensional topological nodal-line semimetal states. Drumhead like surface states protected by the coexistence of time-reversal and mirror symmetry emerge within the two-dimensional regions of the surface Brillouin zone connecting the nodal lines. When SOC is included, unexpectedly, each nodal line evolves into two Weyl nodes (W1, W2) in this centrosymmetric material. Berry curvature calculations show that these nodes occur in a pair and act as either a source or sink of Berry flux. The material also has unique surface states in the form of Fermi arcs, which unlike other known Weyl semimetal, form closed loops of surface states on the Fermi surface. Our theoretical realization of topologically non-trivial states in a superconducting material paves the way towards unraveling the interconnection between topological physics and superconductivity.
\end{abstract}

% abstract OK --ravindra 12/12/10.12

\maketitle

\section{Introduction}%First-level heading:\protect\\ The line
%break was forced \lowercase{via} \textbackslash\textbackslash}

Topological insulators with their non-trivial band topology exhibiting symmetry protected metallic edge/surface states in insulating bulk 2D/3D materials have attracted broad interest in recent times
 \cite{2010ReviewMP-Hasan-colloquium,2011ReviewMP-Qi-TI}. Furthermore, the discovery of these unique non-trivial band topology in semimetals and metals has significantly expanded the family of topological materials much beyond topological insulators \cite{2011PRB-Wan-Iridates,1929ZPhys-Weyl-electron-graviton,2016Naturecommun-Bian-PbTaSe2,2015PRX-Weng-TMmonophosphides-weyl,2015Science-Xu-weyl,2012PRB-Wang-A3Bi-dirac,2011PRB-Burkov-nodalsemimetal,2014naturecommun-Neupane-diracsemimetalCd3As2,2011PRB-Wan-Iridates,2014Naturematerials-Liang-Cd3As2,chen2015nanostructured}. These topological semimetallic/metallic (TSM) materials have unique Fermi surfaces ranging from 0D (points), 1D (lines/loops) to 2D (planes). Till now, four types of topological semimetals/metals has been proposed: (i) 3D Dirac Semimetal (DSM), where 4-fold degenerate bands touch close to the Fermi energy, with their low energy excitations as Dirac fermions, (ii) Weyl Semimetal (WSM), where 2-fold bands touch in the close vicinity of the Fermi level giving gapless node (Weyl Point (WP)), which always comes in pairs of opposite chirality acting as either a source or a sink of the Berry curvature, (iii) Topological nodal line semimetal/metal (TNLSM), where band touching occurs along one-dimensional curves (lines/loops) in $k$ space of materials, and (iv) Topological materials beyond Dirac and Weyl fermions, where linear and quadratic three-, six-, eight-band crossings are stabilized by space group symmetries with spin-orbit coupling and time-reversal symmetries\cite{2016-science-bernevig-beyond-weyl,2016-PRL-Kane-double-dirac,2017-ncomms-Felser-multiple-dirac}. The presence of these unique band topology in TSM materials is constrained by crystallographic symmetries and provide new opportunities to explore exotic topological physics. The distant topological features of these materials are directly reflected in their peculiar electronic/magnetic properties such as large magnetoresistance and ultrahigh mobility \cite{2015Naturephysics-Shekhar-NbP-weyl,2014Naturematerials-Liang-Cd3As2,2014naturecommun-Neupane-diracsemimetalCd3As2}, chiral anomaly and negative magnetoresistance \cite{2012PRB-Zyuzin-weyl,2013PRB-Son-weyl,2015-Nat-Commn-Li-Giant}, and Aharonov-Bohm oscillations \cite{2016-Nat-Commn-Wang-Aharonov}.

Apart from the broad potential applications of these topological materials, the research on the topological quantum matter has garnered much attention due to its intimate connection to high energy physics, by the introduction of Dirac, Weyl, and Majorana fermions (MFs) into the electronic spectra of crystals. Dirac and Weyl's fermionic systems have been realized experimentally, however, experimental realization of MFs is still a challenge. The MFs are predicted to emerge in the topological superconductor, and introducing superconductivity into the topological materials can serve as a host for these fermionic particles \cite{2010ReviewMP-Hasan-colloquium,2011ReviewMP-Qi-TI,2014Science-Grover-supersymmetry,2015PhysicaC-Sasaki-superconductorTI,2014Prl-Yang-DiracWeyl-superconductors,2014Naturephysics-Xu-superconductor}. Superconductivity has been observed in some topological materials with doping \cite{2010Prl-Hor-SupercondcutorCuCi2Se3,2013Prl-Sato-superconductorSnInTe,2014PRB-Zhong-superconductorPbSnTe} and external applied pressure \cite{2016Naturecommun-Qi-superconductorMoTe2}. However, an intrinsic superconducting topological material has not been discovered yet. A discovery of intrinsic superconducting topological material is of utmost importance, and a realization of the same will help in developing more insights into MFs and applications in the future quantum technologies \cite{2008Prl-Fu-superconductivity}. 

Here, using the first-principles density functional theory (DFT) calculations, we have studied the electronic properties of a recently discovered superconducting material CaSn$_3$ with a measured T$_c$ $\approx$ 4.2 K \cite{2015JMCC-Luo-superconductorCaSn3}. The material is found to be a topological nodal line semimetal with drumhead surface states in the absence of spin-orbit coupling (SOC). The bands touch along a line in a large energy window of $\sim$ (-0.5 -- 1.8) eV. Moreover, the material unexpectedly turns into a Weyl semimetal with the inclusion of SOC. The Weyl nodes W1 and W2 are found to occur at $k$ ($\frac{2\pi}{a}$) = (0.2, 0.2, 0.2) and (0.167, 0.167, 0.167), respectively. The Weyl points act like "magnetic" monopoles in momentum space with a charge given by their chirality. They are either a source or sink of "Berry flux," which are confirmed by Berry curvature calculations. A total of 8 pairs of WPs were identified in the Brillouin zone. We also observed the clear Fermi arcs connecting the Weyl nodes in the surface state calculations. This behavior is similar to the many other topological materials \cite{2015PRX-Weng-TMmonophosphides-weyl}, where the WSM phase appears from nodal rings on an inclusion of SOC. Thus, in principle, on tuning the SOC strength one can achieve a transition between the two topological phases. Our prediction of topological states in a superconducting material CaSn$_3$ provides a promising platform for realizing a topologically superconducting material and the synergy between topological physics and superconductivity.

% introduction OK -- ravindra 12/12/10:40

\section{Methodology}

Theoretical calculations were performed within the first-principles density functional theory (DFT) \cite{1965-PhysRev-Kohn-Vxc}  using the Vienna Ab initio Simulation Package (VASP) \cite{1966CMS-Kresse-Ecalcs,1996PRB-Kresse-iterativeEcalcs}. Ion-electron interactions were represented by all-electron projector augmented wave potentials \cite{1994PRB-Blochl-PAW,1999PRB-Kresse-PAW}. The Perdew-Burke-Ernzerhof \cite{1996Prl-Perdew-GGA} generalized gradient approximation (GGA) were used to account for the electronic exchange and correlation. The wave functions were expanded in a plane wave with a high energy cut-off of 500 eV and a Monkhorst-Pack \cite{1976PRB-Monkhorst-BZintegration} \textbf{k}-grid of 15$\times$15$\times$15. The structure was fully relaxed by employing a conjugate gradient scheme until the Hellmann-Feynman forces on the atoms were less than 0.001 eV/\AA{}. Electronic structure calculations were done both with and without SOC. The Berry curvature was calculated by the maximally-localized Wannier functions (MLWF) as implemented in the Wannier90 \cite{2014ComputPhysCommun-Mostofi-Wannier90}. The surface spectrum including the surface projected band structure and Fermi surface were calculated based on the iterative Green's function method \cite{1985JournalofPhysics-Sancho-Greenfunctions} after obtaining the tight-binding Hamiltonian from the MLWF \cite{2014ComputPhysCommun-Mostofi-Wannier90} as implemented in the WannierTools software\cite{opensource-QuanSheng-Wannier-tools}.

% methodology OK -- ravindra 12/12/10:50

\section{Results and Discussion}

\begin{figure}[!ht]
\includegraphics[width=0.8\columnwidth]{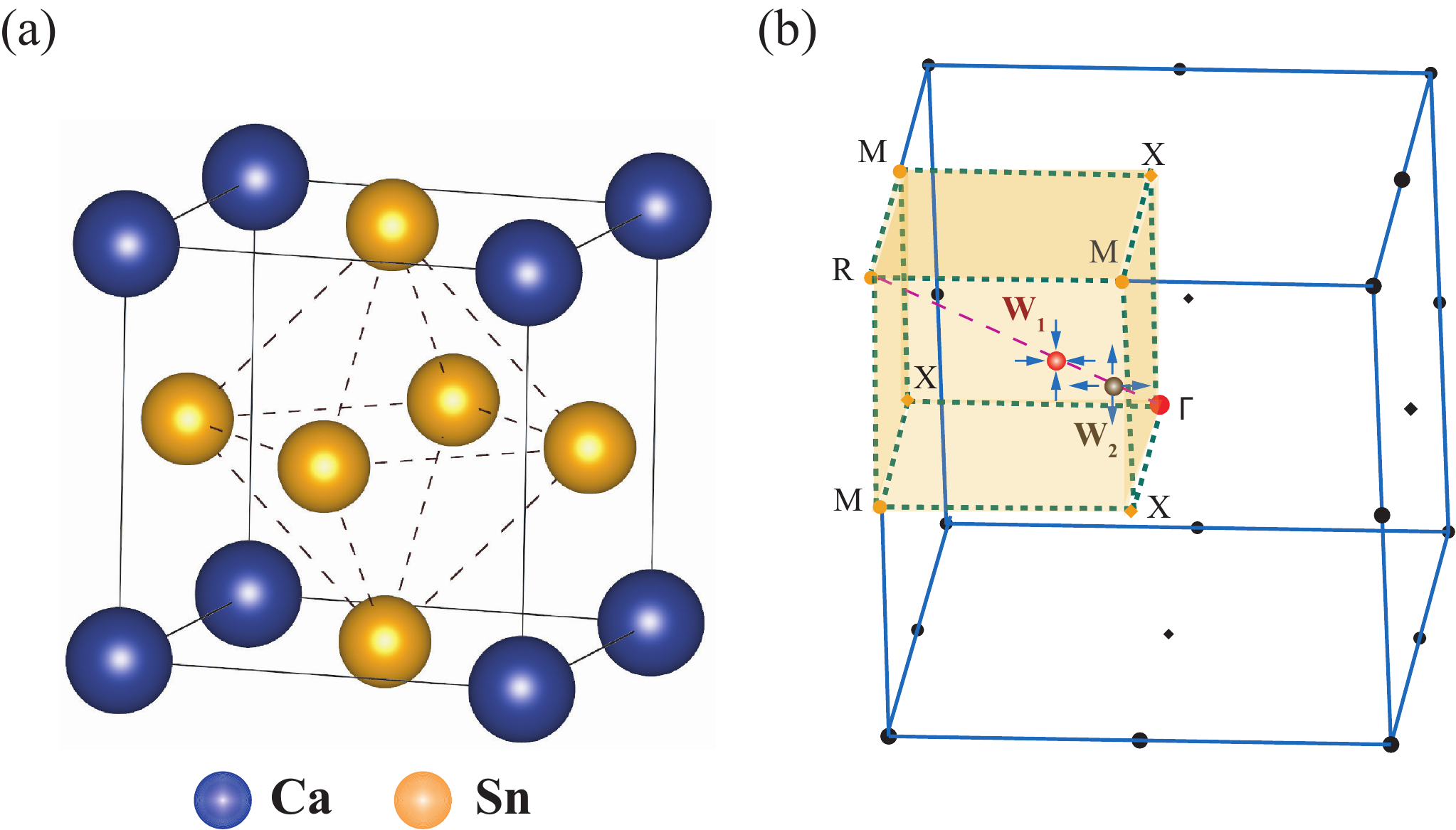}
\caption{(a) Crystal structure of CaSn$_3$ and (b) Brillouin zone of CaSn$_3$ showing the high symmetry points as well as the location of Weyl nodes (W$_1$ and W$_2$). The Weyl nodes act as either sink (W$_1$) and source (W$_2$) of Berry curvature.}
\label{fig:1}
\end{figure}

CaSn$_3$ crystallizes in a cubic AuCu$_3$ type structure with space group Pm$\bar{3}$m \cite{2015JMCC-Luo-superconductorCaSn3}. Ca forms a simple cubic lattice with Sn atoms octahedrally coordinated to each other with no atoms in the center, as shown in Fig. \ref{fig:1}(a). The lattice constant a = 4.74 \AA{} was taken from a recent experimental report on CaSn$_3$ \cite{2015JMCC-Luo-superconductorCaSn3}. The ions were then relaxed fixing the volume as well as the lattice constant of the unit cell.

\begin{figure}[!ht]
\includegraphics[width=\columnwidth]{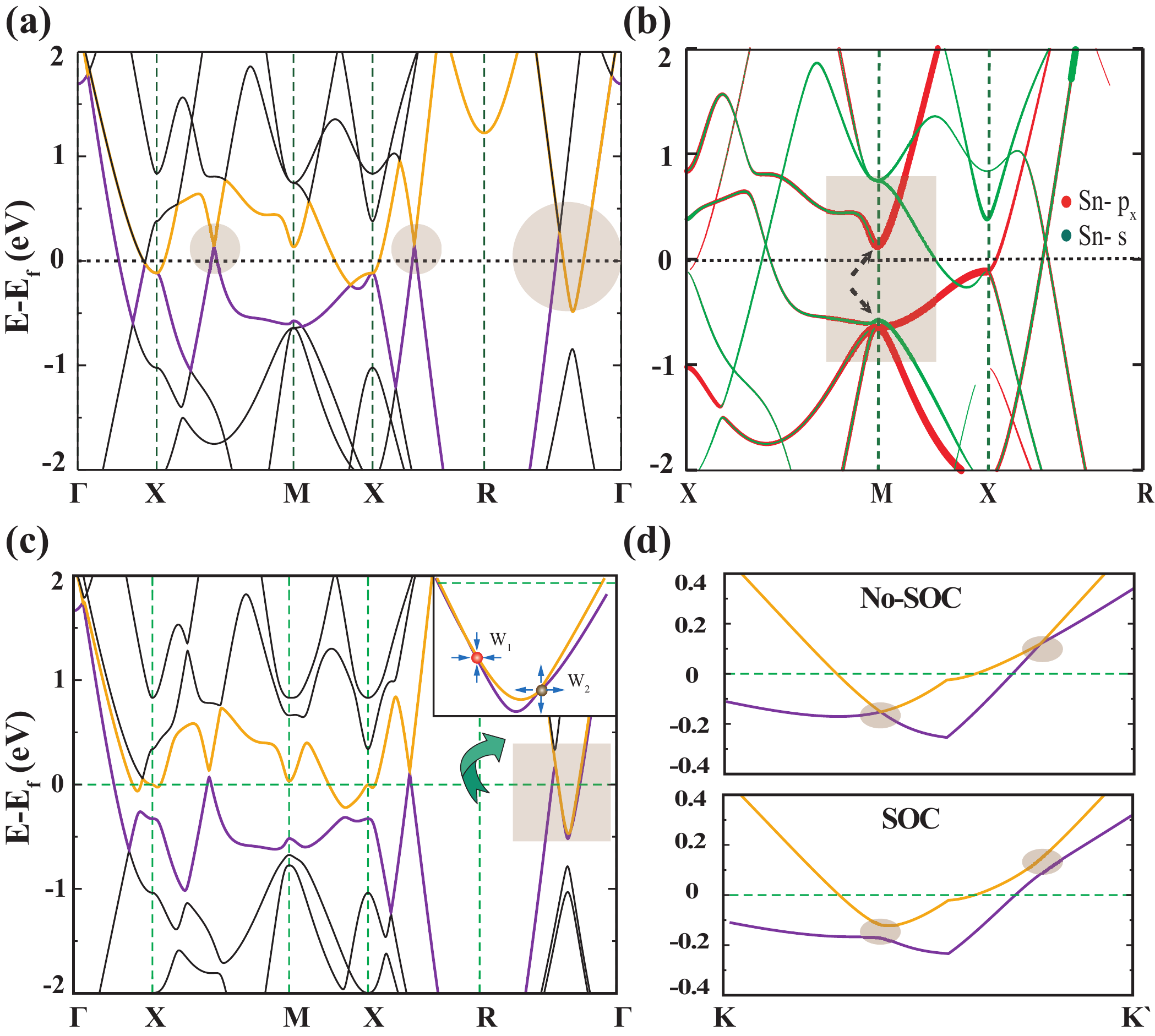}
\caption{(a) Bulk band structure of CaSn$_3$ without spin-orbit coupling (SOC). The shaded area depicts band touching/overlapping regions. (b) Orbital decomposed band structure of CaSn$_3$ showing contribution of Sn $s$ and $p_{x}$ orbitals. The shaded area at M point depicts the band inversion region. (c) Bulk band structure of CaSn$_3$ with SOC. The inset shows the magnified region where the overlapping bands split into two Weyl nodes (W$_1$ and W$_2$). The arrows around the Weyl nodes depict the sign of calculated Berry curvature. (d) Band structure of CaSn$_3$ along a line perpendicular to the R-$\Gamma$ direction showing band touching Dirac cones along the nodal line both with and without SOC.}
\label{fig:2}
\end{figure}

We first study the electronic structure of CaSn$_3$ in the absence of SOC. The band structure along high symmetry lines are shown in Fig. \ref{fig:2}(a). The valence (N) band and conduction (N+1) band has been distinguished with different colors, where N is the total number of valence electrons. It can be seen that there is a band touching (BT) close to the Fermi line along X-M direction as well as X-R direction between N and N+1 bands. To check the orbital character of the bands at these special points as well as at other points along the high symmetry lines, we plotted orbital decomposed band structure as shown in Fig. \ref{fig:2}(b). Only Sn $s$ and $p_x$ orbitals contribution has been shown because of their dominance near the Fermi level. The contribution of $s$ and $p_x$ character are mixed near the BT points; however, a clear band inversion can be seen at the M point indicating possible non-trivial band topology. The valence band near the Fermi level has $s$ character while the conduction band has $p_{x}$ character. This is opposite to the natural order of band filling and clearly, signifies a band inversion. The band inversion at the M point is not due to SOC. Such peculiar band inversion has also been observed in topological nodal line semimetals (AX$_2$ (A = Ca, Sr, Ba; X = Si, Ge, Sn) \cite{2016PRB-Huang-Nodaline-semimetals}) and Cu$_3$PdN \cite{2015Prl-Yu-Nodaline-semimetalCu3PdN}) and a topological semimetal (KNa$_2$Bi) \cite{2016scireports-Sklydneva-KNa2Bi}, indicating the presence of either reflection or point group symmetries in the case of non-centrosymmetric or centrosymmetric materials, respectively \cite{2016PRB-Chan-Ca3P2-drumhead-states}.

An interesting band topology is exemplified in the band structure of CaSn$_3$. The N and N+1 bands overlap between a portion of the R-$\Gamma$ direction. The overlap ranges from $\sim$1.8 eV above the Fermi energy ($E_F$) to $\sim$0.5 eV below $E_F$ (Fig. \ref{fig:2}(a)). On plotting the band structure along a symmetry direction perpendicular to the R-$\Gamma$ direction, we see linearly dispersing Dirac cones near the BT points (Fig. \ref{fig:2}(d)). These Dirac cones persist over the R-$\Gamma$ direction from $k$ ($\frac{2\pi}{a}$) = (0.225, 0.225, 0,225) $\rightarrow$ (0.035, 0.035, 0.035), signifying the material to be topological nodal line semimetal. The presence of special symmetries of the crystal has an important role in deciding the topological protections of nodal lines in a material. It has been proposed that a material should possess either a product of a mirror and time-reversal symmetry or reflection symmetry for the topological protection of nodal lines \cite{2016PRB-Chan-Ca3P2-drumhead-states,2015PRB-Weng-Nodaline-3Dgraphene}. In CaSn$_3$, the topological nodal lines are robust and topologically protected because of the coexistence of both time-reversal and mirror symmetry in the crystal. Due to the special point group symmetry of the crystal,  a total eight nodal lines occur between R-$\Gamma$ in the 3D Brillouin zone of CaSn$_3$. This material, being a metal, the crossings between bands other than N and N+1 are very important. However, in this case, those are at much higher/lower energies to be of any practical significance.

% results and discussion OK upto this part--- Ravindra 12/12/12:45

\begin{figure*}[!ht]
\includegraphics[width=0.9\textwidth]{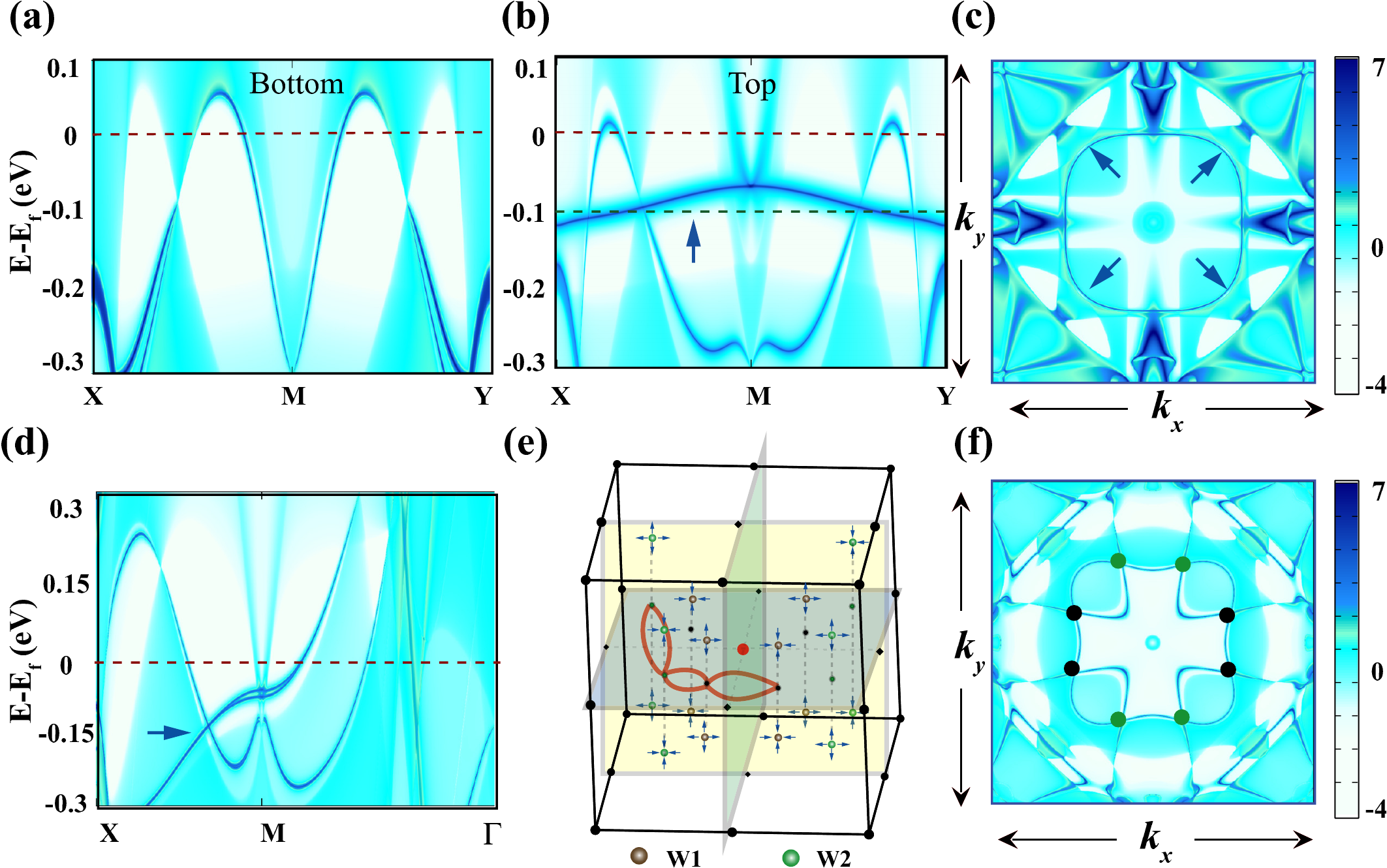}
\caption{Calculated (001) Bloch spectral function without SOC for (a) bottom and (b) top surface. The dark blue lines depict the surface states contribution. The surface states are more dispersive on the bottom surface than that of the top. (c) Fermi surface with the chemical potential at -0.1 eV (green dashed line in (b)). The dark blue regions depict the surface state contribution with the blue arrows pointing towards the drumhead states. (d) Calculated (001) Bloch spectral function with SOC. The blue arrow points toward the surface states with a linearly dispersing Dirac cone. (e) A schematic of the positions of all the Weyl points (W1 and W2) in the Brillouin zone with their chirality as well as their projections on the (001) surface (blue plane). The red loops depict the surface Fermi arcs connecting the Weyl nodes of opposite chirality. (f) (001) Fermi surface of the material with SOC. The dark blue regions depict the surface states. The black (W1) and green (W2) dots denote the Weyl points. These Weyl nodes are connected by surface Fermi arcs, which uniquely form a closed loop in CaSn$_3$.}
\label{fig:3}
\end{figure*}

% results and discussion OK upto this part--- Ravindra 12/12/13:00

The surface states in topological materials are usually related to a topological invariant given by the Berry phase, and a non-zero value guarantees the existence of the topological surface state. In topological nodal line semimetals, it has been shown that the Berry phase is non-zero for any closed path joining the nodal lines \cite{2016PRB-Chan-Ca3P2-drumhead-states,2016PRB-Bian-TlTaSe2drumhead,1993PRB-Vanderbilt-polarization}. This implies that the surface states occur within the two-dimensional regions of the surface Brillouin zone connecting the nodal lines. These surface states have a peculiar feature of a "drumhead" when projected on a 2D plane and a surface flat band when viewed along a 1D line \cite{2016PRB-Huang-Nodaline-semimetals,2011PRB-Burkov-nodalsemimetal,2015PRB-Weng-Nodaline-3Dgraphene}. Thus, to probe the existence of these peculiar surface states, we calculated the surface projected band structure, and Fermi surface on a semi-infinite (001) surface (see Fig. \ref{fig:3}(a),(b)), using the iterative Green's function method \cite{1985JournalofPhysics-Sancho-Greenfunctions} as implemented in WannierTools\cite{opensource-QuanSheng-Wannier-tools}. The white and light-blue regions represent the spectral weight of some additional bulk bands near the surface region, whereas the dark blue curves show the surface states. The Dirac cone arising from the bulk states is observed both in the bottom and top surface along the high symmetry direction X-M-Y. Moreover, there exist surface flat bands connecting the two Dirac cones inside the nodal line ring, namely the "drumhead" states. The surface bands are not flat, as reported in the literature \cite{2015Prl-Yu-Nodaline-semimetalCu3PdN,2015PRB-Weng-Nodaline-3Dgraphene} and are dispersive. The dispersive nature of these drumhead states arises because of the nodal line not being at the same energy level due to the particle-hole asymmetry \cite{2011PRB-Burkov-nodalsemimetal,2015Prl-Yu-Nodaline-semimetalCu3PdN,2015PRB-Weng-Nodaline-3Dgraphene}. These surface flat bands should be directly reflected in the drumhead surface states. Thus, to visualize these drumhead states, we calculated the Fermi surface of the (001) surface (Fig. \ref{fig:3}(c)). The states were calculated at a chemical potential of -0.1 eV. A prominent "drumhead" surface state can be clearly seen in the Fermi surface plot, Fig. \ref{fig:3}(c), signifying the existence of topological nodal lines.% with drumhead surface states in CaSn$_3$.

% Results and Discussion OK -- ravindra 12/12/14:15

In the presence of SOC, the nodal lines are expected to undergo a transition to different topological phases. In many 3D systems, SOC splits the BT points and creates a gap leading to a topological insulator \cite{2015JPSJ-Yamakage-CaAgX-Dirac}. Also if some special symmetries are present nodal line semimetals will persist even in the presence of SOC \cite{2016PRB-Bian-TlTaSe2drumhead}. However, in CaSn$_3$ with SOC the nodal lines evolve into a pair of Weyl nodes, though the material preserves inversion and time-reversal symmetry. The Weyl nodes W1 occur at ($k$ ($\frac{2\pi}{a}$) = 0.2,0.2,0.2), 189 meV below $E_F$, while W2 occur at ($k$ ($\frac{2\pi}{a}$) = 0.167,0.167,0.167), 419 meV below $E_F$ as shown in Fig. \ref{fig:2}(c). These Weyl nodes are tilted and are of type-II character \cite{xu2015structured,soluyanov2015type}. The corresponding Chern numbers are listed in the Table \ref{table:wcc}. The Weyl points act like "magnetic" monopoles in momentum space with their charge represented by the chirality. They are sources of the "Berry flux" similar to the magnetic flux \cite{2011PRB-Wan-Iridates}. Berry flux/curvature is defined as ${F}{=}\nabla_{k}{\times}{A}$, where ${A(k)}=\sum_{n=1}^{N}<{u_{nk}}\mid\nabla_{k}\mid{u_{nk}}>$ is the Berry connection and N is the number of occupied bands. Moreover, the total integral of the Berry curvature (Chern number) of a closed torus around a Weyl point (by Stokes theorem) equals the total "monopole charge" enclosed inside \cite{2011PRB-Wan-Iridates,2015PRX-Weng-TMmonophosphides-weyl}. The topological stability of the Weyl node can be related to the Gauss law, which requires invariance of the total flux inside a given surface. Therefore, Weyl node can only vanish when it annihilates with another Weyl node with opposite chirality and should always come in pairs. The net charge of all the Weyl nodes inside a Brillouin zone should be zero \cite{1981PLB-Nielsen-chiralfermions-theorem}. Thus, to obtain chirality, the Berry curvature was calculated in the vicinity of Weyl nodes. Our calculations indeed show that each node has either positive or negative sign of Berry curvature, thereby acting as either source or sink (Fig. \ref{fig:2}(c), \ref{fig:3}(e)). The symmetry of the crystal engenders eight pairs of Weyl nodes in the 3D Brillouin zone of CaSn$_3$. Furthermore, with SOC, the linear dispersion resembling the character of a Dirac semimetal around the BT point between the X-M and X-R directions (Fig. \ref{fig:2}(a)) develops gaps of magnitude 60 meV and 10 meV respectively. Thus, these BT points are not topologically protected and are accidental band touching.
%%% 

 \begin{table}
 \begin{tabular}{c c c c}\hline \hline
 Weyls& Coordinates & Chern & E -- E$_{F}$\\
 Points & (k$_{x}$, k$_{y}$, k$_{z}$) 2$\pi$/a & number & (meV)$_{F}$\\\hline
 W$_1$ & (0.20, 0.20, 0.20) & --1 & 189 \\\hline
 W$_2$ & (0.167, 0.167, 0.167) & +1 & 419 \\\hline
 \end{tabular}
 \caption{The coordinates, Chern numbers and energies relative to the Fermi level of WPs of CaSn3. Other WPs can be obtained by  symmetry operations $\mathcal{I}$, $\mathcal{C}$2$_{y}$ and $\mathcal{C}$2$_{z}$.}
 \label{table:wcc}
 \end{table}

Another peculiar feature of a Weyl semimetal is the presence of surface Fermi arcs. They must start and end at the projection of two (or more) Weyl points with different "monopole charge," in the surface Brillouin zone. To see the existence of Fermi arcs, we calculated the (001) surface states in the presence of SOC (see Fig. \ref{fig:3}(d), \ref{fig:3}(f)). The black and green circles denote the Weyl points, the light blue and white regions represent the spectral weight of some additional bulk bands near the surface region, whereas the dark blue curves show the surface states. The surface bandstructure shows band crossing surface states with linearly dispersing Dirac cone (Fig. \ref{fig:3}(d)). Moreover, Fermi arcs (Fig. \ref{fig:3}(f)), appear in the surface spectrum connecting the Weyl nodes. The connectivity pattern of Fermi arcs in CaSn$_3$ is similar to the Weyl nodes observed in TaAs\cite{2015Naturecommun-Huang-TaAs-Weyl}, where a pair of Fermi arcs connects the Weyl nodes, instead of a just one.
This peculiarity in the Fermi surface spectrum is observed because of the Weyl nodes of same chiral charge from both top and bottom half of Brillouin zone project at the same point on the (001) plane (Fig. \ref{fig:3}(e)), thereby forming a closed loop of surface states. Moreover, the Weyl nodes with different chirality should be connected by Fermi arcs irrespective of whether they form a pair or not, as evident in this case also.% Thereby, we see that the Weyl nodes W1 and W2 are also connected by Fermi arcs and a closed loop in the Fermi surface is a result of these Fermi arcs connection.

CaSn$_3$ is similar to other materials from TaAs family \cite{2015PRX-Weng-TMmonophosphides-weyl} exhibiting nodal line to Weyl semimetal evolution with SOC.
Moreover, the SOC strength can be diminished in CaSn$_3$ by doping or substituting lighter elements such as Si or Ge in place of Sn. This will potentially allow a topological phase transition between nodal-line semimetal and Weyl semimetal. The details of this topological phase transition and interesting physics arising due to the synergy between the two phases with the tuning of SOC will be discussed in the future work. Moreover, materials hosting Weyl nodes has been found to show large negative magnetoresistance due to a chiral anomaly effect \cite{2012PRB-Zyuzin-weyl,2013PRB-Son-weyl,2015-Nat-Commn-Li-Giant}. Thus, magnetoresistance experiments can be done to verify our predictions in CaSn$_3$. However, the existence of Weyl nodes in centrosymmetric and time-reversal invariant superconducting CaSn$_3$ is intriguing and remains an open question.

\section{Conclusion}
In summary, we have theoretically predicted a non-trivial band topology in a superconducting material - CaSn$_3$. The material has topological line nodes with drumhead surface states in the absence of SOC. The presence of SOC drives the material to Weyl semimetallic state with surface Fermi arcs. The surface Fermi arcs are peculiar in CaSn$_3$, compared to other known Weyl Semimetals, these form closed loop of surface states on the Fermi surface.	
Moreover, unlike previously proposed topological superconductors, the material is intrinsically superconducting and does not require any external tuning such as doping and external pressure to drive superconductivity. In light of these unique properties, the material is expected to serve as an ideal platform to understand the coupling between topological physics and superconductivity.

\begin{acknowledgments}
The authors thank Dr. Swaminathan Venkataraman for valuable discussions. R.S. acknowledges Science and Engineering Research Board, DST, India for a fellowship (PDF/2015/000466). This work is partly supported by the U.S. Army Contract FA5209-16-P-0090. We also acknowledge Materials Research Center and Supercomputer Education and Research Centre, Indian Institute of Science for providing the required computational facilities.  
\end{acknowledgments}

\end{document}